\documentclass[review]{elsarticle}

\pdfoutput=1

\usepackage{url}
\usepackage{breakurl}
\usepackage[breaklinks]{hyperref}

\usepackage{lineno,hyperref}
\usepackage{amsmath}
\usepackage{lscape}
\usepackage{rotating}

\modulolinenumbers[1]

\journal{Scientific African}

\begin{document}

\begin{frontmatter}

\title{A study of COVID-19 data from African countries}

%
%
%
%
%

%
%

\author[add1]{K\'et\'evi A. Assamagan\corref{cor1}}
\ead{ketevi@bnl.gov}
\author[add2]{Somi\'ealo  Azote\corref{cor1}}
\ead{somialo.azote@aims-senegal.org}
\author[add3]{Simon H. Connell}
\author[add4]{Cyrille E. Haliya}
\author[add5]{Toivo S. Mabote}
\author[add6]{Kondwani C. C. Mwale}
\author[add7]{Ebode F. Onyie}
\author[add8]{George Zimba}

\cortext[cor1]{Corresponding Authors}
\address[add1]{Brookhaven National Laboratory, Physics Department, Upton, New York, USA}
\address[add2]{Universit\'e de Lom\'e, D\'epartement de Physique, Lom\'e, Togo}
\address[add3]{University of Johannesburg, Johannesburg, South Africa}
\address[add4]{University of Abomey-Calavi, International Chair in Mathematical Physics and Applications, Cotonou, Benin}
\address[add5]{Universidade Eduardo Mondlane, Grupo de Astrofísica e Ci\^{e}ncias Espaciais, Maputo, Mozambique}
\address[add6]{University of Rwanda, African Center of Excellence for Innovative Teaching and Learning Mathematics and Science, Kigali, Rwanda}
\address[add7]{University of Yaounde I,  Department of Physics,Yaounde, Cameroon}
\address[add8]{University of Jyv\"{a}skyl\"{a}, Department of Physics, Jyv\"{a}skyl\"{a}, Finland}


\begin{abstract}
COVID-19 is a new pandemic disease that is affecting almost every country with a negative impact on social life and economic activities. The number of infected and deceased patients continues to increase globally.  Mathematical models can help in developing better strategies to contain a pandemic. Considering multiple measures taken by African governments and challenging socio-economic factors, simple models cannot fit the data.  We studied the dynamical evolution of COVID-19 in selected African countries.  We estimated a time-dependent reproduction number, $R_0$ for each country studied to offer further insights into the spread of COVID-19 in Africa. We found that at the beginning the pandemic, $R_0 \leq 4$ for all the countries studied; three months later, $R_0 \sim 1$ with fluctuations in-between.
\end{abstract}


\end{frontmatter}


\section{Introduction}
\label{sec:into}

\par\noindent COVID-19 has spread to the entire world within a few months~\cite{Worldh2020}. The World Health Organization (WHO) predicts that 29 to 44 million Africans could be infected with SARS-CoV-2 during the first year of the pandemic and 83 to 190 thousand Africans could die if they don't uphold containment measures~\cite{worldhth2020,WHO2020}. This grim prediction suggests that most African countries have a lower transmission rate than the other regions of the world such as Europe, the United States of America, and China~\cite{worldhth2020}. However, the low transmission rate may prolong the outbreak over several years, putting pressure on economic resources. Most African countries are struggling because of lack of essential medical resources such as test kits, personal protective equipments and ventilators. The containment measures such as frequent hand washing, isolation, contact tracing, and social distance are a challenge in Africa---around 60\% of the African population lives below the poverty line~\cite{WorldBank} and cannot afford the basic hygienic amenities. The densely populated slums of Africa make social distancing impossible and burdens the isolation centers.  In Africa, the outbreak of COVID-19 has already claimed thousands of lives, rendered millions jobless, increased insecurity and poverty level.  A number of studies have been performed on the evolution and impact of COVID-19 in Africa, and on the African responses to the pandemic~\cite{Massinga2020,martinez2020, Ebrahim2020, Etkind2020,Adegboye2020,MIF2020,AU2020,Lone2020}.

Models for pandemics are necessary for understanding the cause, source, spread, and planning outbreak containment \cite{Van2020,Zhao2020,Abila2020,Bendaif2020,medinilla2020, dahab2020,McBryde2020,lauer2020,meehan2020,asamoah2020}. The simplest of these models is the SIR model; it describes disease transmission and propagation in three categories,  namely the susceptible, infected and recovered fractions of a population~\cite{sinkala2020}.  An improved version of the SIR model is the SEIR model which proposes four stages: susceptible, exposed, infectious, and removed population densities~\cite{nachega2020}. Simple models for COVID-19 do not offer reliable insights or predictions to inform African policymakers~\cite{sinkala2020}. The models become complex when one includes more socio-economic factors. One such model is the SIDARTHE~\cite{Giordano2020} which considers eight stages of epidemic evolution.

In this paper,  we analyzed COVID-19 data from Benin, Mozambique, Rwanda, Togo and Zambia. We tested the SIDARTHE model on these data and estimated basic reproduction numbers. This may improve our understanding of the spread of COVID-19 in Africa, although the numbers of tests are small relative to the sizes of the populations.  We offer suggestions to keep the basic reproduction number below one, to slow and contain the spread. In Section~\ref{sec:mod}, we present the mathematical model used in the studies reported in this paper. In Section~\ref{sec:stra}, we discuss the analysis strategy and present the results. In Section~\ref{sec:disc}, we discuss the implications of the results, and we offer concluding remarks in Section~\ref{sec:conc} 

%
%

\section{Model}
\label{sec:mod}

\begin{figure}[!h]
\centering
\includegraphics[width=1.0\textwidth]{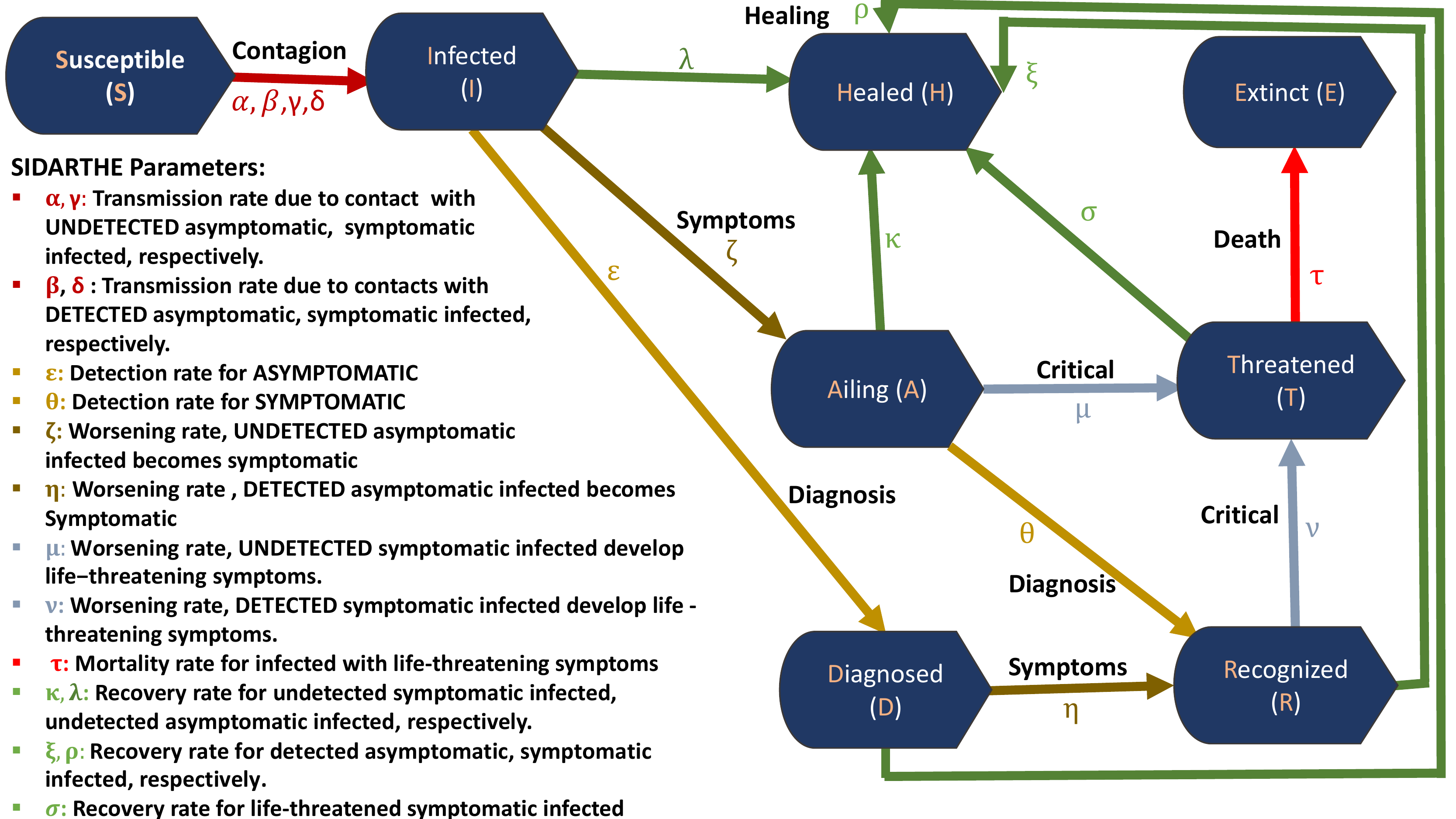}
\caption{\emph{Graphical representation of different compartments of the SIDARTHE model \cite{Giordano2020}: S stands for susceptible, the total population of the case study region or country; I, infected (asymptomatic infected undetected); D, diagnosed (asymptomatic infected  detected); A,
ailing (symptomatic infected undetected); R, recognized (symptomatic infected, detected); T, threatened (infected with life-threatening symptoms, detected); H, healed (recovered); E, extinct (dead).} }
\label{eq:Fig1}
\end{figure}

\noindent To have confidence in a model, one needs suitable fits to existing data and verifiable predictions.  Here, we describe the SIDARTHE dynamical model, developed to study the spread of COVID-19 in Italy~\cite{Giordano2020}. The strength of this model comes from the fact that it considers the various measures taken by Italian government to contain the disease~\cite{Giordano2020}.  It is a mean-field epidemiological model with eight time-dependent compartments, namely Susceptible, Infected, Diagnosed, Ailing, Recognized, Threatened, Healed and Extinct categories, as shown in Figure~\ref{eq:Fig1}. This model describes the dynamic spread of the disease when  social distancing, lockdown, testing, contact tracing, treatment, curfew, and/or quarantine are implemented as containment strategies in a population. 

The following mathematical system of eight differential equations describes the SIDARTHE model~\cite{Giordano2020}:

\begin{equation}
\label{equ:dif}
\begin{array}{lcl}
\frac{\mathrm{d}S(t)}{\mathrm{d}t} &= & -S(t)\left( \alpha I(t)+\beta D(t) +\gamma A(t) +\delta R(t)\right), 	
	 \\
	\frac{\mathrm{d}I(t)}{\mathrm{d}t} &=& S(t)\left( \alpha I(t)+\beta D(t) +\gamma A(t) +\delta R(t)\right) -\left( \lambda +\varepsilon+\zeta\right) I(t),\\
	\frac{\mathrm{d}D(t)}{\mathrm{d}t} &=&\varepsilon I(t)-\left(\eta+\rho \right) D(t),\\
	\frac{\mathrm{d}A(t)}{\mathrm{d}t} &=&\zeta I(t)-(\theta+\mu+\kappa)A(t),\\
	\frac{\mathrm{d}R(t)}{\mathrm{d}t} &=&\eta D(t)+\theta A(t)-(\nu +\xi)R(t),\\
	\frac{\mathrm{d}T(t)}{\mathrm{d}t} &=&\mu A(t)+\nu R(t)-(\sigma + \tau)T(t),\\
	\frac{\mathrm{d}H(t)}{\mathrm{d}t} &=&\lambda I(t) + \rho D(t) +\kappa A(t)+ \xi R(t) + \sigma T(t), \\
	\frac{\mathrm{d}E(t)}{\mathrm{d}t} &=&\tau T(t).	
\end{array}
\end{equation}
The basic reproduction number, $R_0$, is an epidemiological parameter to describe the contagiousness or transmissibility of infections~\cite{Giordano2020}. Biological, socio-economic, environmental and behavioral factors affect $R_0$. It is a parameter used to study the dynamics of an infectious disease. An outbreak ends if $R_0<1$ and continues if $R_0>1$.  $R_0$ indicates of the potential magnitude of an outbreak, and can be used to estimate the fraction of the population to be vaccinated to stop the spread. However, because of its complex dependence on many factors, $R_0$ is often modeled and, as a result, depends on model parameters and assumptions. Therefore, one must apply $R_0$ with great caution. The SIDARTHE model defines $R_0$ as follows~\cite{Giordano2020}:
\begin{equation}
\label{equ:r0}
R_0 = \frac{\alpha}{r_1}+
\frac{\beta\times\epsilon}{r_1 \times r_2}+
\frac{\gamma\times\zeta}{r_1 \times r_3}+
\frac{\delta\times\eta\times\epsilon}{r_1 \times r_2 \times r_4}+
\frac{\delta\times\zeta\times\theta}{r_1 \times r_3 \times r_4},
\end{equation}
with
\begin{eqnarray}
r_1 &=& \epsilon+\zeta+\lambda, \nonumber \\
r_2 &=& \eta+\rho, \nonumber \\
r_3 &=& \theta+\mu+\kappa, \\
r_4 &=& \nu+\xi, \nonumber \\
r_5 &=& \sigma+\tau. \nonumber
\end{eqnarray}
We adapted the SIDARTHE model to consider the containment measures taken by African countries and the impact of socio-economic conditions in Africa. In Section~\ref{sec:stra}, we discuss the analyses of data from Benin, Mozambique, Rwanda, Togo and Zambia, and the application of the SIDARTHE model to these data. 
\section{Analysis}
\label{sec:stra}

\noindent We collected the first three months of the official data on COVID-19 from Benin, Mozambique, Rwanda, Togo and Zambia. We got the data from the official website of each country. One team member who is a resident (or is a native) of a country was in charge to compile and follow the measures taken. The same team member was also responsible to understand the tests performed in that country. The data came in categories of active, recovered, dead and total cases. Compared to the SIDARTHE stages of pandemic evolution, it is straightforward to establish the following associations: the recovered cases correspond to the healed compartment and the dead cases to extinct category shown in Figure~\ref{eq:Fig1}. The active cases do not have a direct correspondence in the model. One needs to understand the tests to define an association of the active cases to the model. From the eight stages in the SIDARTHE, the active cases in the data should, at the bare minimum, map to the sum of the recognized and threatened categories. However, depending on whether asymptomatic or ailing persons were tested and counted, the active cases might contain some of them. To compare data to the model, we defined the active cases as the sum of the recognized, threatened and ailing (or diagnosed) compartments---this is not an exact correspondence because of the complexity of the testing and counting procedures.  In addition, the total cases also do not map directly to any stage of the model. In the data, the total cases are the sum of the active, recovered and dead cases. In the model, we built the total cases as the sum of the model active cases  and healed and extinct compartments shown in Figure~\ref{eq:Fig1}.

After we defined the mapping of the data onto the model compartments or stages, we matched the model to the data by adjusting the model parameters depending on whether the active, recovered and/or dead cases were increasing or decreasing. We solved the eight differential equations in Eq.~(\ref{equ:dif}) by Euler discretization to estimate the parameters from best match between model and data. Subsequently, we computed $R_0$ according to Eq.~(\ref{equ:r0}). The result is an extraction of a time-dependent $R_0$ from the estimated parameters. Depending on the evolution of the pandemic and the responses measures imposed, the parameters in Equations~\ref{equ:dif} will change with time; therefore, $R_0$, shown in Equation~\ref{equ:r0}, will also change with time, as illustrated in the bottom plots of Figures~\ref{eq:Fig2}--\ref{eq:Fig6}.

In the following subsections, we discuss each country, one-by-one.

\subsection{Case of Benin}

\begin{figure}[!h]
\centering
\includegraphics[width=1.0\textwidth]{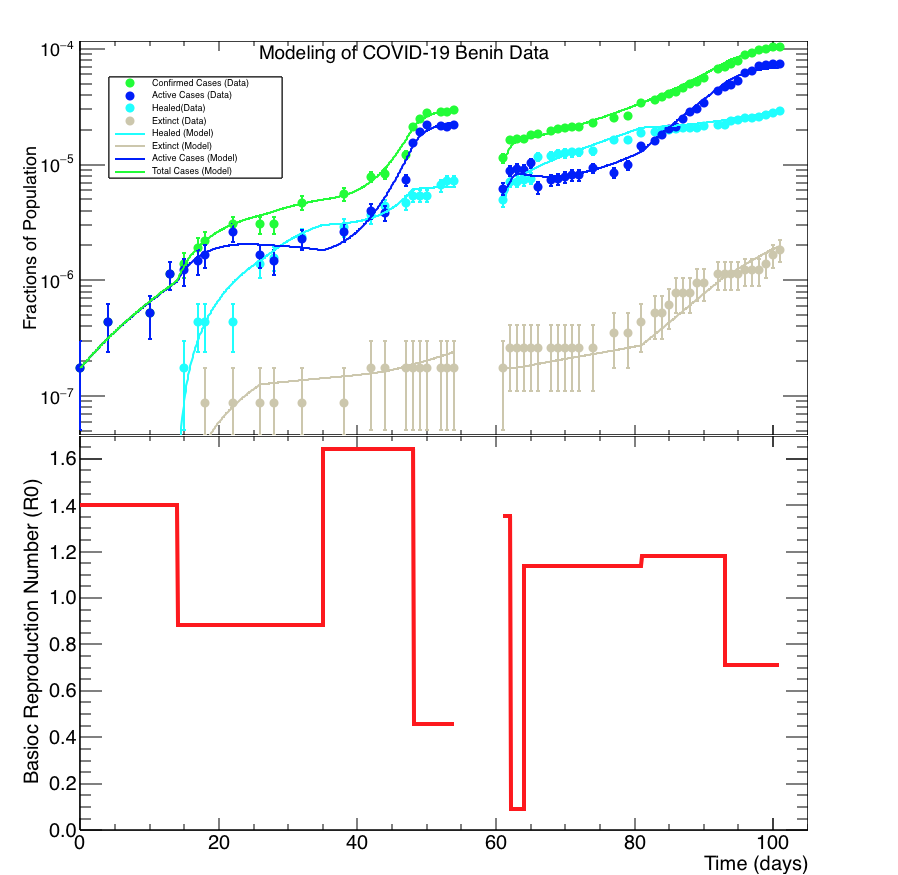}
\caption{\emph{In the top plot, we show the official data compiled by the government of Benin. Day 0 is March 16, 2020. The uncertainties shown on the data points are statistical only. We normalized the data to a population of  11.5 million. Superimposed is the SIDARTHE model applied to the data. The bottom plot shows the resulting $R_0$ for Benin as a function of time.} }
\label{eq:Fig2}
\end{figure}

\noindent They identified the first case on March 16, 2020, and the government took immediate containment measures such as limitation in border crossings, compulsory quarantine of people entering the country by air, suspension of government and business missions outside the country, suspension of all demonstrations and non-essential sporting, cultural, religious or political events, closure of mosques and churches, social distance, hygiene and wearing of masks requirements. From March 30 to May 11, 2020, schools and universities were closed. They imposed a total lockdown on the regions---Cotonou, Abomey-Calavi, Allada, Ouidah, Sèmè-Podji, Porto-Novo, Akpro-Missérété and Adjarra---most exposed to the pandemic. The government engaged in an awareness campaign through the media and the police force. They  encouraged people to inform the authorities about anyone who returned to the country and did not self-isolate. From May 11, the government lifted the lockdown of the aforementioned regions and by June 2, and activities resumed with mandatory social distance and the wearing of masks. We collected the official data compiled by the government and modeled it as shown in Figure~\ref{eq:Fig2} where one sees that there is a period between Day 54 and Day 61 where they posted no data. 
In the top panel of Figure~\ref{eq:Fig2}, there is a systematic shift in the data before Day 54 compared to after Day 61. This is because of the difference in the reporting of the test results. Before May 19, the government reported results of both the rapid diagnostic and polymerase chain reaction (PCR) tests. After May 19, following the WHO guidelines, the government started reporting only the PCR test results, although they continued to perform the rapid diagnostic tests. We see a good match between the SIDARTHE model and the data. The bottom panel of  Figure~\ref{eq:Fig2} shows the resulting time-dependent $R_0$ from the modeling. Table~\ref{table:benin} shows the model parameters that best match the data of Benin. We find that the basic reproduction number rarely exceed two; however, it fluctuates. After May 19, $R_0$ rarely exceeds one because only the PCR test results were being reported; however, it may also be because of the effectiveness of the measures implemented by the government.

\subsection{Case of Mozambique}
\noindent In Mozambique, they detected the first case on March 22, 2020. The individual was a Mozambican national who had traveled to the United Kingdom. The patient showed mild symptoms. The health authorities placed him in isolation at home and under clinical supervision. The government closed schools and universities on March 23, suspended the issuance of entry visas, and cancelled the ones already issued. They also suspended social events with over 50 people. They required travelers to self-quarantine.
\begin{figure}[!h]
\centering 
\includegraphics[width=1.0\textwidth]{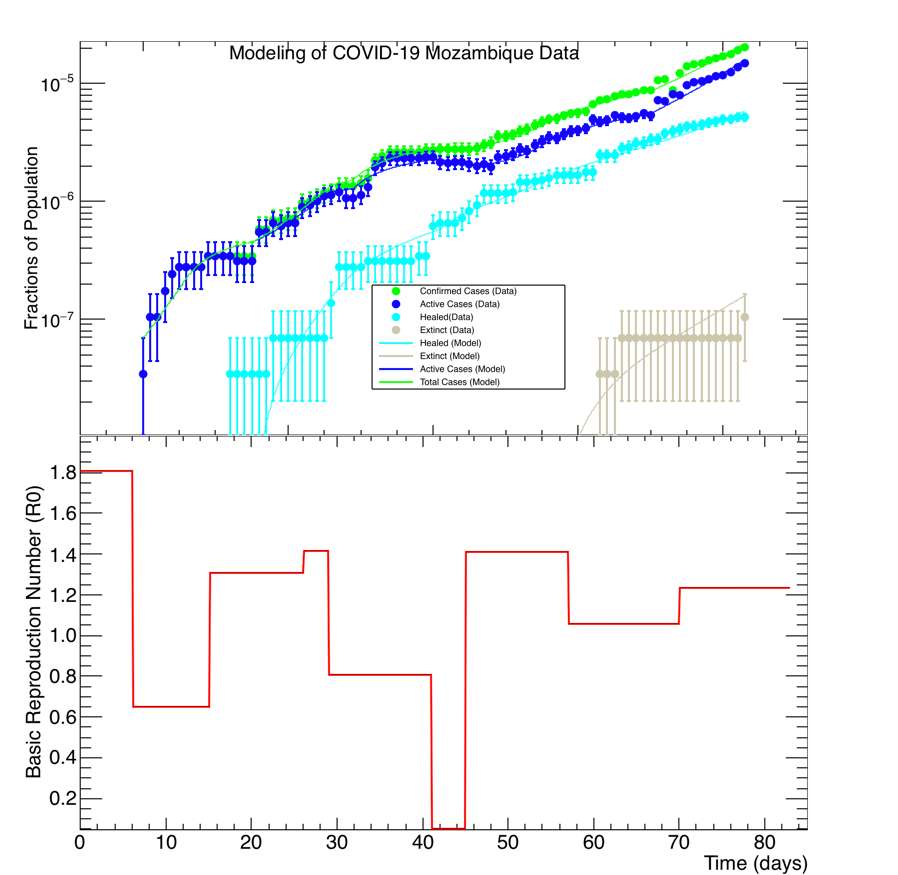}
\caption{\emph{In the top plot, we show the official data compiled by the government of Mozambique, normalized to a population of 29 million. The uncertainties shown on the data points are statistical. Superimposed is the SIDARTHE model applied to the data. Day 0 is March 22, 2020. The bottom plot shows the resulting $R_0$ for Mozambique as a function of time.} }
\label{eq:Fig3}
\end{figure}
The country went into a state of emergency on April 1. They extended the state of emergency successively: on April 29 until May 30; then until the end of June; on June 28 until July 29. On May 12, they suspended international flights until May 30, except for humanitarian, cargo or state flights. However, they did not impose a lockdown. At the time of writing this article, the government and local authorities were studying schools re-opening strategies. Figure~\ref{eq:Fig3} shows the COVID-19 data of Mozambique with the modeling of the SIDARTHE; in the top panel, we see good agreement between the model and the data for all the cases of the dead, recovered and active fractions of the population. As a result, the total cumulative cases are also well modeled. In the bottom panel of Figure~\ref{eq:Fig3}, we show the extracted $R_0$ which remains below two for the entire period shown. The $R_0$ for Mozambique fluctuates. Between Day 40 and Day 45, it dropped significantly. After Day 45, it stays slightly above one. Table~\ref{table:mozambique} shows the model parameters that best match the data of Mozambique.

\subsection{Case of Rwanda}

\noindent On March 14, 2020, Rwanda confirmed its first case of COVID-19. It was a foreign national who arrived in the country on March 8. The individual showed no symptoms upon arrival; however, he reported to a health facility on March 13 and tested positive. They started testing symptomatic cases right away, before they identified the first case. Contact tracing and testing of asymptomatic cases started on March 14. From March 15, they postponed schools, religious activities, weddings until further notice and implemented social distance measures.  Because of an increase in the number of cases, the authorities took additional safety measures on March 21: they imposed a lockdown by closing of bars, boarders, airports and markets, except for those selling food and hygienic essentials. They required masks in all public places and provided markets and shops with sanitizers.   
\begin{figure}[!h]
\centering
\includegraphics[width=1.0\textwidth]{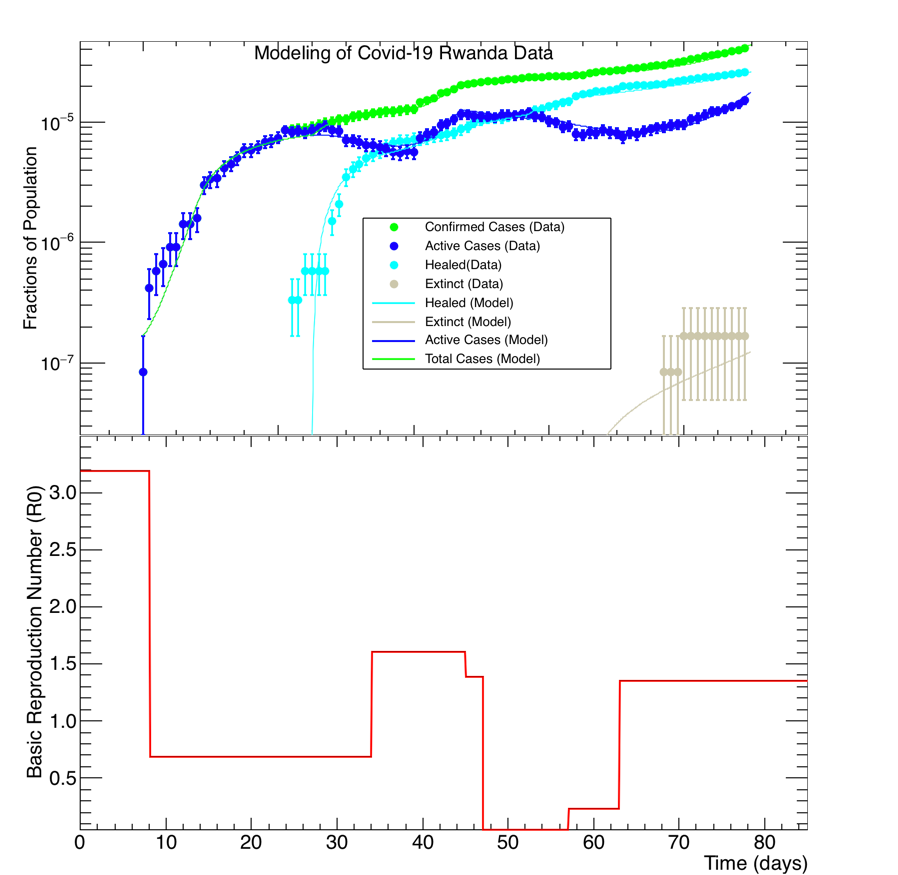}
\caption{\emph{In the top plot, we show the official data compiled by the government of Rwanda, normalized to a population of 12 million. The uncertainties shown are statistical. Day 0 is March 14, 2020. Superimposed on the data is the SIDARTHE model applied to the data. The bottom plot shows the resulting $R_0$ for Rwanda as a function of time.} }
\label{eq:Fig4}
\end{figure}
Figure~\ref{eq:Fig4} shows the Rwanda COVID-19 data on the top panel; we superimpose the modeling of the data and see  good agreement in the dead, recovered and active cases. As a result, the total cases are also well modeled. From the modeling, we derived $R_0$ for Rwanda as shown in the bottom panel of Figure~\ref{eq:Fig4}. The initial $R_0$ is above three, but drops well below one after about a week because of the swift reaction of the government and the public. After a few weeks, the $R_0$ rose above one, most likely because of the difficulties to observe the measures imposed. We see another reduction in $R_0$ around Day 47; around Day 64, it went up to about 1.5. Table~\ref{table:rwanda} shows the model parameters that best match the data of Rwanda.

\subsection{Case of Togo}

\noindent Togo recorded its first case of COVID-19 on March 6, 2020; the individual was a Togolese national who had traveled abroad. The government implemented containment measures right away, such as contact tracing, monitoring of persons under quarantine, testing of symptomatic cases, and surveillance at points of entry, borders and airports.
\begin{figure}[!h]
\centering
\includegraphics[width=1.0\textwidth]{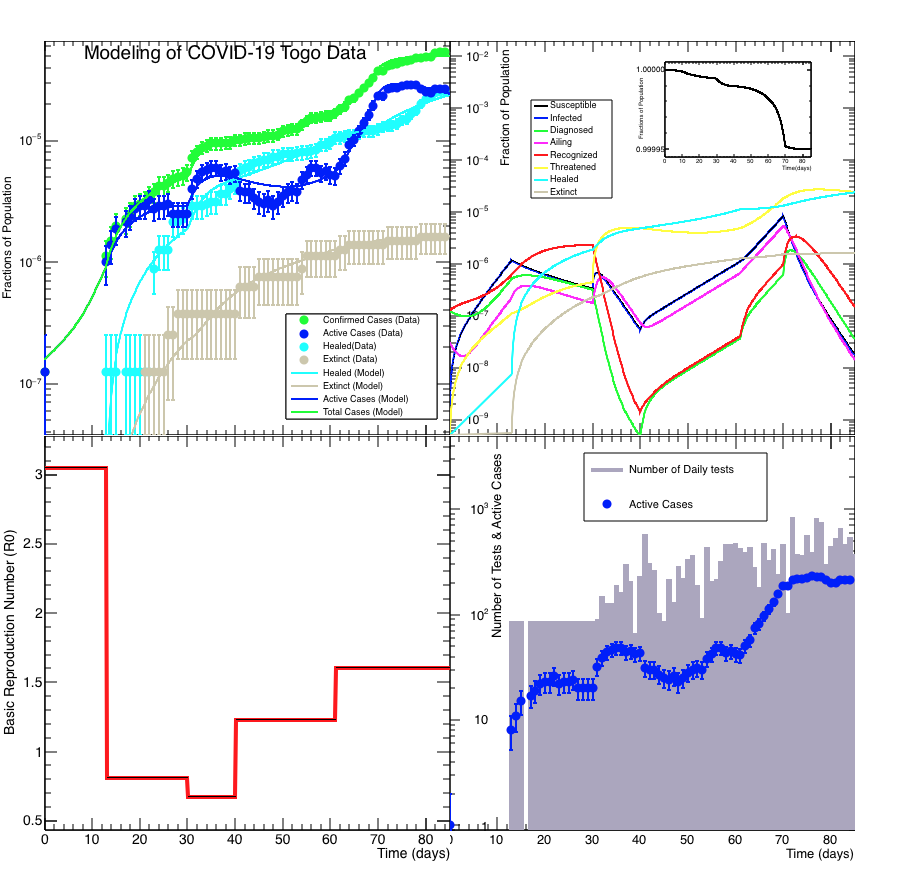}
\caption{\emph{In the top plot left, we show the official data compiled by the government of Togo. The uncertainties shown are statistical. Day 0 is March 6, 2020. Superimposed is the SIDARTHE model applied to the data. The top right plot shows the details of the SIDARTHE model for Togo with the time evolution of the eight stages of the pandemic. The embedded picture in the top right panel shows the distribution of the susceptible population. We normalized the top plots to a population of 8 million. The bottom left plot shows the $R_0$ for Togo as a function of time. The bottom right plot shows the number of active cases superimposed onto the number of the daily tests done in Togo.} }
\label{eq:Fig5}
\end{figure}
 After an extraordinary meeting of the council of ministers on March 16, the government established the following measures: suspension flights from Italy, France, Germany, and Spain; cancellation of all international events for three weeks; self-isolation of people coming from high-risk countries; border closure; and prohibition of events with over 100 people effective from March 19. For at least two-and-a-half months, schools, universities, churches, saloons, bars, etc., were closed. They imposed a curfew from 9:00pm to 6:00am. They tested truck drivers crossing the borders; then they allowed the trucks to proceed to their destinations under surveillance. If the drivers had been in contact with confirmed cases, they placed them under quarantine. On April 7, the government started massive tests of both symptomatic and asymptomatic persons in cities with over ten cases. From June 9, they lifted the curfew. However, the government made the wearing of masks compulsory; also, they required hand washing before access to public or private services or markets. 

We used the containment measures to tune the model parameters as a function of time.  Figure~\ref{eq:Fig5} shows the data and the model;  on the top left panel, we see good agreement in the dead and recovered cases. For the active cases, the agreement is good in the earlier and later time periods. The mis-modeling observed in the middle time period is likely related to the difficulty in defining accurately the active cases in the model as mentioned in Section~\ref{sec:stra}. For the total cases, the model agrees with the data in the entire period shown. In the top right panel, we show the time evolutions of all the eight stages of the SIDARTHE model for Togo. The bottom left panel of Figure~\ref{eq:Fig5} shows the $R_0$ for Togo. We see that in first 2 weeks, $R_0$ was about three. It dropped in the subsequent few weeks because of the effectiveness of the containment measures and the social awareness campaign. However, after Day 40, the $R_0$ rose; this is because between May 5--20, the number of cases sharply increased when neighboring countries re-opened their borders. This led to an influx of imported cases from Togolese nationals that returned to Togo. The bottom right plot shows the number of daily tests and the active cases---the same active cases shown on the top left plot. The active cases show structures the distribution where, periodically, the cases increased or dropped. To model the data accurately, we tried to understand whether these structures were correlated with the number of daily tests or related to the dynamical evolution of the pandemic. As shown in the bottom right plot of Figure~\ref{eq:Fig5}, we found no corrections between the number of daily tests and the active cases.  Until Day 30, Togo reported only the total number of tests done, not the daily test numbers. In the bottom right panel of Figure~\ref{eq:Fig5}, we see a flat distribution up to Day 30—we took an average by dividing the total number tests over the number of days. After Day 30, the histogram in the bottom right plot of Figure~\ref{eq:Fig5} shows the reported daily test numbers. Table~\ref{table:togo} shows the model parameters that best match the data of Togo.

\subsection{Case of Zambia}

\noindent On March 18, 2020, Zambia reported its first two cases of COVID-19. Zambia hosts the Southern Africa Regional Collaborating Center of the Africa CDC (Center for Disease Control) and  has  been  coordinating  the  response  at  the regional level. The government has put in place a contingency plan that outlines the country's preparedness. The  government  continues  to  enforce  the  measures  and  interventions  to  control  the  spread  countrywide. The public health safety measures implemented include the closure of schools and higher learning institutions; wearing of a mask while out in public; continued screening of travelers into Zambia; redirection of all international flights to land and depart from Kenneth Kaunda International Airport only; suspension of non-essential travels to countries with confirmed COVID-19 cases; restriction of public gatherings; restaurants to operate only on take away and delivery basis; and closure of all bars, nightclubs, cinemas, gyms and casinos. On May 8, the control measures were further reviewed: restaurants may revert to their normal operation; cinemas, gyms and casinos may also reopen; they made an appeal to proprietors of hotels, lodges, tour operators, event management companies and others---who  voluntarily  closed  their  business  to  ensure  the  safety  of  their  staff  and  clientele---to  consider reopening; bars and taverns remained closed pending further review of the measures, depending on the evolution of the pandemic; they allowed only examination classes in primary and secondary schools to reopen. The first classes reopened on June 1 with  enforced  public  health  guidelines  in  place: the  reopening of  business  premises  and  schools is  subject  to  adherence  to  public  health  regulations, guidelines  and  certifications. The government continues to update response activities on a regular basis.
\begin{figure}[!h]
\centering
\includegraphics[width=1.0\textwidth]{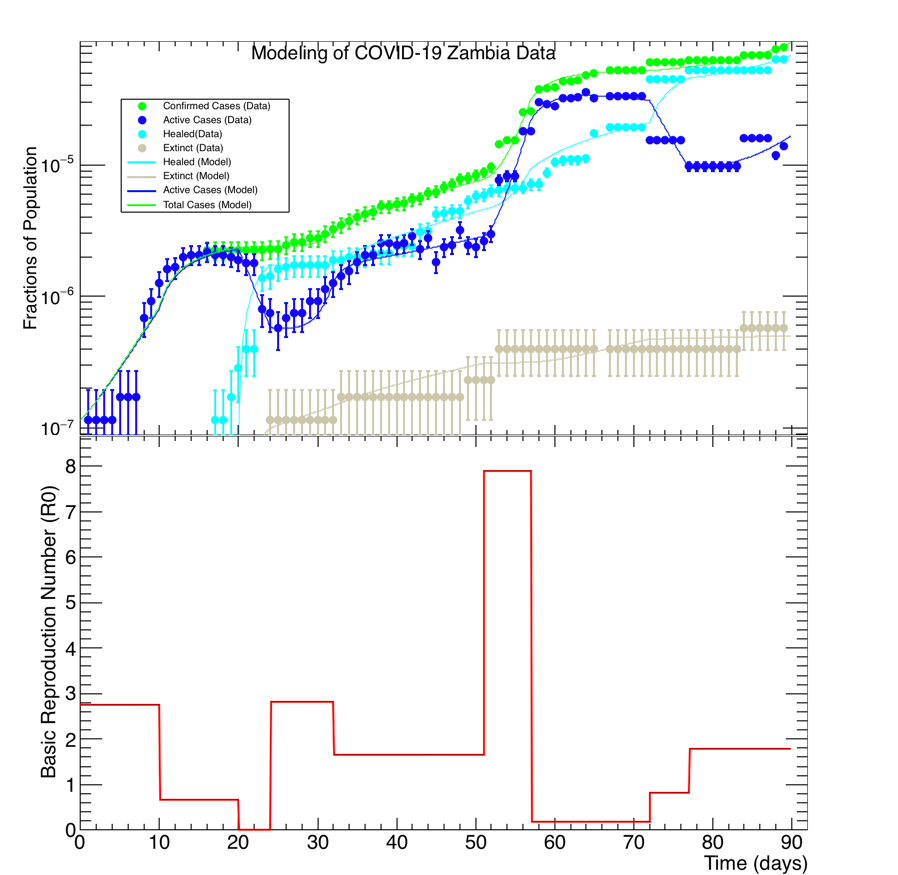}
\caption{\emph{In the top plot, we show the official data compiled by the government of Zambia, normalized to a population of 17.5 million. We show only statistical uncertainties. Day 0 is March 18, 2020. Superimposed is the SIDARTHE model applied to the data. The bottom plot shows the resulting $R_0$ for Zambia as a function of time.} }
\label{eq:Fig6}
\end{figure}

Figure~\ref{eq:Fig6} shows the COVID-19 data of Zambia and its SIDARTHE modeling.  The death rate and the total cases are well modeled. The trends of the recovered and active cases are fairly well modeled. The $R_0$ for Zambia started close to three but dropped below one within a few weeks. It rose again, and around Day 50 it rose to about eight until Day 55. This is  because of a significant increase in the reported numbers of daily cases around Day 50.  On May 8, the government dispatched a team of health workers to Nakonde---a town next to Tanzania---to provide technical support and enhance port health services, community surveillance and disinfection of public places. They tested truck drivers, community members, health care workers, staff of lodges and the Immigration Department. The prior number of total cases was 167 and on May 9, they had 85 cases, almost a 50\% increase. Seventy-six of the 85 cases were from Nakonde. Between May 9 and 16, they reported high daily cases of 174 and 208. One hundred twenty-six of the 174 cases were from Nakonde and  196 of the 208  cases were also from Nakonde. These increases in the daily cases, concentrated around Nakonde, explain the high $R_0$ in Day 50-55. The $R_0$ dropped again around Day 55 until about Day 70 when it increased above one. Table~\ref{table:zambia} shows the model parameters that best match the data of Zambia.

\section{Discussion}
\label{sec:disc}

\noindent For the all the countries studied, $R_0$ started above one with a few imported cases. Within a few weeks, $R_0$ dropped below one because of the swift and decisive reactions of the governments and the awareness campaigns. The people reacted well initially and followed the authorities' directives. Unfortunately, $R_0$ did not stay below one for a long period; in all the cases studied, the basic reproduction number rose again above one after a few weeks---because of difficulties in adhering to the measures when the people face other socio-economic challenges. The rise of $R_0$ after it had fallen initially may also because of complacency, fake news, and misinformation---some believe that COVID-19 is a scam, Africans are immune, and/or the disease has no impact in tropical climates, because of the low transmission rate mentioned in Section~\ref{sec:into}. That the initial responses were effective to bring $R_0$ below one is an encouragement that African countries can contain the spread. The challenge is to maintain the containment measures long enough to bring $R_0$ permanently below one. A continuous campaign of community engagement with regular briefings is important; so are an active combat against fake news and misinformation. They should maintain the lockdown and social distance measures notwithstanding the socio-economic adversities. Economic relief is necessary for the people with hardships exacerbated by these measures; this will motivate adherence to the containment plans and that will stop the pandemic~\cite{otu2020tackling, mehtar2020limiting, ihekweazu2020africa, rosenthal2020covid}. 

A comment on the studies reported in this paper is their validity, given the numbers of limited tests performed. We show in Figure~\ref{eq:Fig5} that the number of cases are not correlated with the limited number of tests. The statistical samples used are significant; therefore, the conclusions are valid.  One may extrapolate these results to the larger populations of the countries studied to determine, for example, the number of people to vaccinate. However,  from the limited tests, we cannot extrapolate to infer the total number of infections in the country. We also caution extrapolating to the future to make predictions; this is because, as we have shown in Figures~\ref{eq:Fig2}--\ref{eq:Fig6}, the basic reproduction number fluctuates. Only detailed modeling from first principles in biology, medicine, physics, epidemiology and sociology may offer a framework for viable predictions. 

\section{Conclusion}
\label{sec:conc}

\noindent We have studied COVID-19 data from Benin, Mozambique, Rwanda, Togo and Zambia. We modeled the data from these countries with the SIDARTHE model, and extracted a time-dependent basic reproduction number for each country studied. Our studies showed that at the onset the pandemic, the basic reproduction numbers, $R_0<4$, for all the countries studied. The initial reactions of African governments and populations were effective to bring the basic reproduction number below one after a few weeks. Three months later, $R_0 \sim 1$, with fluctuations in between---relaxation and difficulties to maintain the measures over time drove the basic reproduction number in a time-dependent cyclic pattern of rises and falls. We suggest that African countries find satisfactory economic supports for their most disadvantaged populations. This will encourage adherence to the containment plans.  	
\section*{Acknowledgements}

\noindent We thank Professor John Ellis (University of London) for useful discussions. Toivo S. Mabote would like to thank Professor Doutor Cl\'audio Mois\'es Paulo (Universidade Eduardo Mondlane) for academic advice and mentorship. We acknowledge support and mentorship from ASP---the African School of Fundamental Physics and Applications. We received no financial support for this work.


%
%
%

\bibliography{mybibfile}

\biboptions{sort&compress}

\newpage

\section*{Supplementary Material}

\renewcommand{\thetable}{SM1}

\begin{table}[!h]
\centering
\begin{tabular}{c|c|c|c|c|c|c|c|c}
\hline\hline
Time (day) &  $\alpha$ & $\beta$ & $\gamma$ & $\delta$ & $\epsilon$ & $\theta$ & $\mu$ & $\nu$ \\
                  &  $\zeta$ & $\eta$ & $\lambda$ & $\rho$ & $\kappa$ & $\chi$ & $\sigma$ & $\tau$ \\ \hline
$0-14$       & 0.28 & 0.10 & 0.28 & 0.10 & 0.25 & 0.37 & 0.10 & 0.20 \\
                  &  0.32 & 0.32 & $0.10\times10^{-2}$ & 0.00 & 0.00 & 0.00 & 0.00 & $0.40\times10^{-2}$ \\ \hline
$15-26$     & 0.60 & 0.40 & 0.60 & 0.40 & 0.97 & 0.97 & 0.97 & 0.97 \\
                  &  0.97 & 0.97 & 0.10 & 0.06 & 0.06 & 0.06 & 0.06 & 0.006 \\ \hline
$27-35$     & 0.65 & 0.65 & 0.65 & 0.65 & 0.02 & 0.03 & 0.50 & 0.50 \\
                  &  0.50 & 0.50 & 0.95 & 0.04 & 0.04 & 0.04 & 0.04 & 0.001 \\ \hline
$36-48$     & 0.70 & 0.70 & 0.70 & 0.70 & 0.02 & 0.02 & 0.60 & 0.60 \\
                  &  0.80 & 0.80 & 0.20 & 0.001 & 0.001 & 0.001 & 0.001 & 0.001 \\ \hline
$49-54$     & 0.25 & 0.10 & 0.25 & 0.10 & 0.06 & 0.08 & 0.75 & 0.75 \\
                  &.  0.75 & 0.75 & 0.02 & 0.001 & 0.001 & 0.001 & 0.001 & 0.50$\times 10^{-2}$ \\ \hline\hline
$61-62$     & 0.85 & 0.80 & 0.85 & 0.80 & 0.06 & 0.08 & 0.45 & 0.45 \\ 
                  & 0.90 & 0.90 & 0.80 & 0.09 & 0.09 & 0.09 & 0.09 & 0.50$\times 10^{-2}$ \\ \hline
$63-64$     & 0.05 & 0.01 & 0.05 & 0.01 & 0.06 & 0.08 & 0.95 & 0.95 \\
                  &  0.95 & 0.95 & 0.001 & 0.08 & 0.08 & 0.08 & 0.08 & 0.50$\times 10^{-2}$ \\ \hline
$65-81$     & 0.50 & 0.10 & 0.50 & 0.10 & 0.06 & 0.08 & 0.75 & 0.75 \\
                  &  0.73 & 0.73 & 0.001 & 0.08 & 0.08 & 0.08 & 0.08 & 0.70$\times 10^{-2}$ \\ \hline
$82-93$     & 0.50 & 0.01 & 0.50 & 0.01 & 0.06 & 0.08 & 0.75 & 0.75 \\
                  &  0.73 & 0.73 & 0.65$\times 10^{-3}$ & 0.01 & 0.01 & 0.01 & 0.01 & 0.28$\times 10^{-2}$ \\ \hline
$\geq 94$  & 0.30 & 0.01 & 0.30 & 0.01 & 0.06 & 0.08 & 0.75 & 0.75 \\
                  &  0.73 & 0.73 & 0.65$\times 10^{-3}$ & 0.01 & 0.01 & 0.01 & 0.01 & 0.002 \\
\hline\hline
\end{tabular}
\caption{The SIDARTHE model parameters (day$^{-1}$) that best match the first few months of COVID-19 data of Benin. The uncertainties in these numbers result from the statistical uncertainties in data as shown in Figure~\ref{eq:Fig2}. Day~0 is March 16, 2020. The relative uncertainties are large---up to $\sim100$\% at the onset of the pandemic, but decrease over time with more data to $\sim10$\%. }
\label{table:benin}
\end{table}

\renewcommand{\thetable}{SM2}
\begin{table}[!h]
\centering
\begin{tabular}{c|c|c|c|c|c|c|c|c}
\hline\hline
Time (day) &  $\alpha$ & $\beta$ & $\gamma$ & $\delta$ & $\epsilon$ & $\theta$ & $\mu$ & $\nu$ \\
                  & $\zeta$ & $\eta$ & $\lambda$ & $\rho$ & $\kappa$ & $\chi$ & $\sigma$ & $\tau$ \\ \hline
$0-8$         & 0.57 & 0.30 & 0.38 & 0.30 & 0.25 & 0.37 & 0.10 & 0.20 \\
                  & 0.32 & 0.32 & 0.10$\times 10^{-2}$ & 0.00 & 0.00 & 0.00 & 0.00 & 0.00 \\ \hline
$9-25$       & 0.20 & 0.004 & 0.10 & 0.004 & 0.08 & 0.37 & 0.10 & 0.20 \\
                  & 0.32 & 0.32 & 0.10$\times 10^{-2}$ & 0.00 & 0.00 & 0.00 & 0.00 & 0.00 \\ \hline
$26-34$    & 0.65 & 0.65 & 0.65 & 0.65 & 0.02 & 0.03 & 0.95 & 0.95 \\
                  & 0.50 & 0.50 & 0.95 & 0.04 & 0.04 & 0.04 & 0.04 & 0.00 \\    \hline
$35-45$    & 0.70 & 0.70 & 0.70 & 0.70 & 0.02 & 0.02 & 0.60 & 0.60  \\
                  & 0.80 & 0.80 & 0.20 & 0.025 & 0.025 & 0.025 & 0.025 & 0.00 \\   \hline
$46-47$     & 0.50 & 0.50 & 0.50 & 0.50 & 0.06 & 0.08 & 0.65 & 0.65 \\
                  &  0.65 & 0.65 & 0.08 & 0.001 & 0.001 & 0.001 & 0.001 & 0.00 \\ \hline 
$48-57$     & 0.001 & 0.80$\times 10^{-3}$ & 0.001  & 0.80$\times 10^{-3}$ & 0.026 & 0.038 & 0.025 & 0.025 \\
                  & 0.035 & 0.035 & 0.0015 & 0.005 & 0.005 & 0.005 & 0.030 & 0.00 \\ \hline   
$58-63$     & 0.15 & 0.10 & 0.15 & 0.10 & 0.26 & 0.38 & 0.65 & 0.65 \\
                  & 0.65 & 0.65 & 0.60 & 0.045 & 0.045 & 0.045 & 0.045 & 0.00 \\ \hline  
$\geq 64$  & 0.42 & 0.42 & 0.42 & 0.42 & 0.07 & 0.09 & 0.55 & 0.55 \\
                  & 0.50 & 0.50 & 0.085 & 0.03 & 0.03 & 0.03 & 0.03 & 0.50$\times 10^{-3}$ \\                                                                                             
\hline\hline
\end{tabular}
\caption{The SIDARTHE model parameters (day$^{-1}$) that best match the first few months of COVID-19 data of Rwanda. The uncertainties in these numbers result from the statistical uncertainties in data as shown in Figure~\ref{eq:Fig3}. Day~0 is March 14, 2020. The relative uncertainties are large---up to $\sim100$\% at the onset of the pandemic, but decrease over time with more data to $\sim10$\%.}
\label{table:rwanda}
\end{table}

\renewcommand{\thetable}{SM3}

\begin{table}[!h]
\centering
\begin{tabular}{c|c|c|c|c|c|c|c|c}
\hline\hline
Time (day) &  $\alpha$ & $\beta$ & $\gamma$ & $\delta$ & $\epsilon$ & $\theta$ & $\mu$ & $\nu$ \\
                      $\zeta$ & $\eta$ & $\lambda$ & $\rho$ & $\kappa$ & $\chi$ & $\sigma$ & $\tau$ \\ \hline
$0-6$         & 0.40 & 0.20 & 0.30 & 0.20 & 0.25 & 0.37 & 0.20 & 0.30 \\ 
                  & 0.32 & 0.32 & 0.10$\times 10^{-3}$ & 0.00 & 0.00 & 0.00 & 0.00 & 0.00 \\ \hline
$7-15$       & 0.20 & 0.004 & 0.10 & 0.004 & 0.08 & 0.37 & 0.20 & 0.30 \\ 
                  & 0.32 & 0.32 & 0.10$\times 10^{-3}$ & 0.00 & 0.00 & 0.00 & 0.00 & 0.00 \\ \hline
$16-27$    & 0.50 & 0.30 & 0.50 & 0.30 & 0.07 & 0.08 & 0.65 & 0.65 \\
                 & 0.60 & 0.60 & 0.10 & 0.004 & 0.004 & 0.004 & 0.004 & 0.00 \\ \hline   
$27-29$    & 0.17 & 0.08 & 0.17 & 0.08 & 0.02 & 0.02 & 0.20 & 0.20 \\
                 & 0.1 & 0.1 & 0.07 & 0.01 & 0.01 & 0.01 & 0.01 & 0.00 \\ \hline
$30-41$    & 0.30 & 0.20 & 0.30 & 0.20 & 0.06 & 0.08 & 0.65 & 0.65 \\
                 & 0.65 & 0.65 & 0.06 & 0.01 & 0.01 & 0.01 & 0.01 & 0.00 \\ \hline
$42-45$    & 0.001 & 0.80$\times 10^{-3}$ & 0.001 &  0.80$\times 10^{-3}$ & 0.026 & 0.038 & 0.025 & 0.025 \\
                 & 0.035 & 0.035 & 0.0015 & 0.005 & 0.005 & 0.005 & 0.03 & 0.00 \\ \hline
$46-57$    & 0.55 & 0.30 & 0.55 & 0.030 & 0.26 & 0.38 & 0.55 & 0.55 \\
                 & 0.55 & 0.55 & 0.08 & 0.02 & 0.02 & 0.02 & 0.02 & 0.00. \\ \hline
$58-70$    & 0.22 & 0.10 & 0.22 & 0.10 & 0.06. & 0.08 & 0.40 & 0.40 \\
                 & 0.30 & 0.30 & 0.015 & 0.02 & 0.02 & 0.02 & 0.02 & 0.001 \\ \hline 
$\geq 71$ & 0.40 & 0.30 & 0.40 & 0.30 & 0.06 & 0.08 & 0.60 & 0.60 \\
                 & 0.60 & 0.60 & 0.015 & 0.02 & 0.02 & 0.02 & 0.02 & 0.001 \\                                                        
\hline\hline
\end{tabular}
\caption{The SIDARTHE model parameters (day$^{-1}$) that best match the first few months of COVID-19 data of Mozambique. The uncertainties in these numbers result from the statistical uncertainties in data as shown in Figure~\ref{eq:Fig4}. Day~0 is March 22, 2020. The relative uncertainties are large---up to $\sim80$\% at the onset of the pandemic, but decrease over time with more data to $\sim10$\%.}
\label{table:mozambique}
\end{table}

\renewcommand{\thetable}{SM4}
\begin{table}[!h]
\centering
\begin{tabular}{c|c|c|c|c|c|c|c|c}
\hline\hline
Time (day) &  $\alpha$ & $\beta$ & $\gamma$ & $\delta$ & $\epsilon$ & $\theta$ & $\mu$ & $\nu$ \\
                  & $\zeta$ & $\eta$ & $\lambda$ & $\rho$ & $\kappa$ & $\chi$ & $\sigma$ & $\tau$ \\ \hline
$0-13$     & 0.48 & 0.025 & 0.38 & 0.025 & 0.171 & 0.37 & 0.017 & 0.027 \\
                  & 0.125 & 0.125 & 0.00 & 0.00 & 0.00 & 0.00 & 0.00 & 0.00 \\ \hline
$14-30$    & 0.22 & 0.004 & 0.12 & 0.004 & 0.16 & 0.37 & 0.012 & 0.022 \\
                  & 0.19 & 0.19 & 0.02 & 0.035 & 0.035 & 0.035 & 0.035 & 0.05 \\ \hline 
$31-40$    & 0.55 & 0.45 & 0.55 & 0.45 & 0.006 & 0.006 & 0.98 & 0.98 \\
                 & 0.97 & 0.97 & 0.65 & 0.02 & 0.02 & 0.02 & 0.02 & 0.008 \\ \hline
$41-61$    & 0.67 & 0.50 & 0.66 & 0.50 & 0.02 & 0.02 & 0.58 & 0.58 \\
                 & 0.45 & 0.45 & 0.50 & 0.025 & 0.025 & 0.025 & 0.025 & 0.008 \\ \hline
$62-70$   & 0.51 & 0.50 & 0.51 & 0.50 & 0.06 & 0.08 & 0.58 & 0.58 \\
                & 0.58 & 0.58 & 0.06 & 0.001 & 0.001 & 0.001 & 0.001 & 0.003 \\ \hline    
$\geq 71$     & 0.03 & 0.03 & 0.03 & 0.03 & 0.26 & 0.38 & 0.41 & 0.42 \\
                & 0.42 & 0.42 & 0.002 & 0.025 & 0.025 & 0.025 & 0.025 & 0.003 \\                                                             
\hline\hline
\end{tabular}
\caption{The SIDARTHE model parameters (day$^{-1}$) that best match the first few months of COVID-19 data of Togo. The uncertainties in these numbers result from the statistical uncertainties in data as shown in Figure~\ref{eq:Fig5}. Day~0 is March 6, 2020. The relative uncertainties are large---up to $\sim80$\% at the onset of the pandemic, but decrease over time with more data to $\sim10$\%.}
\label{table:togo}
\end{table}

\renewcommand{\thetable}{SM5}
\begin{table}[!h]
\centering
\begin{tabular}{c|c|c|c|c|c|c|c|c}
\hline\hline
Time (day) &  $\alpha$ & $\beta$ & $\gamma$ & $\delta$ & $\epsilon$ & $\theta$ & $\mu$ & $\nu$ \\
                  &  $\zeta$ & $\eta$ & $\lambda$ & $\rho$ & $\kappa$ & $\chi$ & $\sigma$ & $\tau$ \\ \hline
$0-10$      & 0.35 & 0.15 & 0.30 & 0.15 & 0.10 & 0.20 & 0.10 & 0.10 \\
                 & 0.30 & 0.30 & 0.02 & 0.00 & 0.00 & 0.00 & 0.00 & 0.00 \\ \hline
$11-20$    & 0.25 & 0.20 & 0.25 & 0.25 & 0.01 & 0.02 & 0.80 & 0.80 \\
                 & 0.70 & 0.70 & 0.005 & 0.002 & 0.002 & 0.002 & 0.00 & 0.00 \\   \hline
$21-24$    & 0.00 & 0.00 & 0.00 & 0.00 & 0.02 & 0.02 & 0.40 & 0.40 \\
                 & 0.40 & 0.40 & 0.80 & 0.32 & 0.32 & 0.32 & 0.32 & 0.02 \\ \hline                 
$25-32$    & 0.95 & 0.75 & 0.95 & 0.75 & 0.312 & 0.62 & 0.55 & 0.55 \\
                 & 0.65 & 0.65 & 0.01 & 0.0025 & 0.0025 & 0.0025 & 0.0025 & 0.01 \\ \hline 
$33-51$   & 0.14 & 0.12 & 0.14 & 0.12 & 0.02 & 0.03 & 0.03 & 0.03 \\
                & 0.50 & 0.50 & 0.33 & 0.025 & 0.025 & 0.025 & 0.025 & 0.01 \\ \hline
$52-57$   & 0.55 & 0.60 & 0.55 & 0.60 & 0.02 & 0.03 & 0.045 & 0.045 \\
                & 0.65 & 0.65 & 0.12 & 0.022 & 0.022 & 0.022 & 0.022 & $0.80\times 10^{-3}$ \\ \hline
$58-72$   & 0.03 & 0.005 & 0.03 & 0.005 & 0.04 & 0.05 & 0.20 & 0.20 \\
                & 0.35 & 0.35 & 0.04 & 0.02 & 0.02 & 0.02 & 0.02 &  $0.50\times 10^{-3}$  \\ \hline
$73-77$   & 0.35 & 0.25 & 0.35 & 0.25 & 0.04 & 0.05 & 0.095 & 0.095 \\
                & 0.25 & 0.25 & 0.48 & 0.25 & 0.25 & 0.25 & 0.25 &  $0.10\times 10^{-3}$  \\ \hline
$\geq 78$ & 0.50 & 0.22 & 0.50 & 0.22 & 0.04 & 0.05 & 0.095 & 0.095 \\
                & 0.45 & 0.45 & 0.65 & 0.035 & 0.035 & 0.035 & 0.035 & $0.20\times 10^{-3}$  \\                                                                                         
\hline\hline
\end{tabular}
\caption{The SIDARTHE model parameters (day$^{-1}$) that best match the first few months of COVID-19 data of Zambia. The uncertainties in these numbers result from the statistical uncertainties in data as shown in Figure~\ref{eq:Fig6}. Day~0 is March 18, 2020. The relative uncertainties are large---up to $\sim80$\% at the onset of the pandemic, but decrease over time with more data to $\sim10$\%.}
\label{table:zambia}
\end{table}

\end{document}